\documentclass[12pt]{article}

\usepackage{amsfonts, amssymb, amsthm}

\newtheorem{tw}{Theorem}

\newtheorem{de}{Definition}

\newtheorem{co}{Corollary}

\newtheorem{lem}{Lemma}

\renewcommand{\theenumi}{ \roman{enumi}}

\newcommand{\be}{\begin{equation}}

\newcommand{\ee}{\end{equation}}

\newcommand{\bea}{\begin{eqnarray}}

\newcommand{\eea}{\end{eqnarray}}

\usepackage{epsfig}
\usepackage{amsfonts, amssymb, amsthm}

\begin{document}

\begin{center}
{\LARGE{\bf{States in the Hilbert space formulation and the phase space formulation  of quantum mechanics }}}
\end{center}

\bigskip\bigskip

\begin{center}
 J. Tosiek\footnote{E-mail address:  tosiek@p.lodz.pl} and P. Brzykcy\footnote{E-mail address:  800289@edu.p.lodz.pl}
\end{center}

\begin{center}

{\sl  Institute of Physics, Technical University of  \L\'{o}d\'{z},\\ W\'{o}lcza\'{n}ska 219, 90-924 \L\'{o}d\'{z}, Poland.}\\
\medskip

\end{center}

\vskip 1.5cm
\centerline{\today}
\vskip 1.5cm

\begin{abstract}
W consider the problem of testing if a given matrix in the Hilbert space formulation of quantum mechanics or a function in the phase space formulation of quantum theory represent a quantum state. We propose several practical criteria to recognise states in these both versions of quantum physics. After minor modifications they can be applied to check positivity of any operators acting in a Hilbert space  or positivity  of any functions from an algebra with a Weyl type $*$ -- product.
\end{abstract}


%%%%%%%%%%%%%%%%%%%%%%%%%%%%%%%%%%%%%%%%%%%%%%%%%%%%%%%%%%%%%%%%%%%%%%%%%%%%%%%%%%%%%%%%%%%%%%%%%%%%%%%%%%%%%%%%%%%%%%%%%%%%%%%%%%%%%%%%%%%%%%%%%%%%%%%%%%%%%%%%%%%%%%%%%%%%%%%%%%%%%%%%%%%%%%%%%%%%%%%%%%%%%%%%%%%%%%%%%%%%%%%%%%%%%%%%%%%%%%%%%%%%%%%%%%%%%%%%%%%%%%%%%%%%%%%%%%%%%%%%%%%%%%%%%%%%%%%%%%%%%%%%%%%%%%%%%%%%%%%%%%%%%%%%%%%%%%%%%%%%%%%%%%%%%%%%%%%%%%%%%%%%%%%%%%%%%%%%%%%%%%%%%%%%%%%%%%%%%%%%%%%%%%%%%%%%%%%%%%%%%%%%%%%%%%%%%%%%%%%%%%%%%%%%%%%%%%%%%%%%%%%%%%%%%%%%%%%%%%%%%%%%%%%%%%%%%%%%%%%%%%%%%%%%%%%%%%%%%%%%%%%%%%%%%%%%%%%%%%%%%%%%%%%%%%%%%%%%%%%%%%%%%%%%%%%%%%%%%%%%%%%%%%%%%%%%%%%%%%%%

\section{Introduction}

\setcounter{de}{0}
\setcounter{tw}{0}
\setcounter{lem}{0}
\setcounter{co}{0}

Mathematical structure of quantum mechanics has been formed from the two ingredients: observables and states.
A notion of an observable is not unique and it depends on a chosen convention. In the Hilbert space formulation of quantum mechanics measured
quantities are represented by linear operators. It is usually assumed that these operators are self --adjoint \cite{bohm}, \cite{prug}. In the phase
space formulation of quantum theory observables are represented by smooth real functions on a symplectic manifold \cite{tat}--\cite{pleban}, but other
functions can also be considered.

On the contrary, a sense of a quantum state is precise. For example, in the Hilbert space formulation of quantum mechanics normalizable pure states are represented by
vectors of the unitary length from a Hilbert space ${\cal H}.$ Thus a set of all states is a convex set spanned by the pure states \cite{zy}.

Formal construction of a space of states in quantum mechanics is based on a $*\;$-- algebra ${\cal A}.$ The algebra ${\cal A}$ contains some subset of
the set of observables. Let the symbol ${\cal F}$ denote a set of linear functionals over the algebra ${\cal A}.$ Quantum states are represented by
positive functionals $f \in {\cal F}$ i.e. such that 
\be
\label{0}
 \forall \, A \in  {\cal A} \; \; \;  f( A^+ A) \geq 0
\ee
and satisfying the normalization condition 
\be
\label{10}
f({\bf 1})=1.
\ee
By ${\bf 1}$ we denote the unity of the algebra ${\cal A}.$ The expected value of an observable $A \in {\cal A}$ in a state $f \in {\cal F}$ equals
\[
\big< A\big> := f(A).
\]
Detailed analysis of the algebraic attempt to quantum states can be seen in  \cite{th}.

In the Hilbert space formulation of quantum theory the algebra  ${\cal A}$ is 
 the algebra $B({\cal H})$ of bounded operators defined on the whole Hilbert space ${\cal H}.$   The functional action is determined by the formula
\[
f(A):= {\rm Tr} (\hat{\varrho} \hat{A}), \;\;\; \hat{A} \in B({\cal H}),
\]
where the symbol $\hat{\varrho}$ denotes a density operator. More information about this topic can be found in \cite{land}.

The natural question arises, how to test whether a given functional represents a state. Definition \ref{def1} as well as geometric one are hardly applicable.
Hence we propose several criteria for solving this problem. 
We consider  the Hilbert space formulation and the phase space formulation of quantum mechanics.
 The most complex task is checking positivity. Our results referring to positivity are quite general and can be used e.g. in construction of a time operator \cite{prz}.

The sign `$*$' plays a dual role in the text. It denotes both: an involution in the algebra ${\cal A}$ and a $*$-- product. A square matrix $C$ of a dimension  ${\dim {\cal H}} \times {\dim {\cal H}}$ is symbolised by $ \Big[ \big<\varphi_i | C| \varphi_j \big> \Big]_1^{\dim {\cal H}}$ while an element of this matrix is represented as $\big<\varphi_i | C| \varphi_j \big>.$

%%%%%%%%%%%%%%%%%%%%%%%%%%%%%%%%%%%%%%%%%%%%%%%%%%%%%%%%%%%%%%%%%%%%%%%%%%%%%%%%%%%%%%%%%%%%%%%%%%%%%%%%%%%%%%%%%%%%%%%%%%%%%%%%%%%%%%%%%%%%%%%%%%%%%%%%%%%%%%%%%%%%%%%%%%%%%%%%%%%%%%%%%%%%%%%%%%%%%%%%%%%%%%%%%%%%%%%%%%%%%%%%%%%%%%%%%%%%%%%%%%%%%%%%%%%%%%%%%%%%%%%%%%%%%%%%%%%%%%%%%%%%%%%%%%%%%%%%%%%%%%%%%%%%%%%%%%%%%%%%%%%%%%%%%%%%%%%%%%%%%%%%%%%%%%%%%%%%%%%%%%%%%%%%%%%%%%%%%%%%%%%%%%%%%%%%%%%%%%%%%%%%%%%%%%%%%%%%%%%%%%%%%%%%%%%%%%%%%%%%%%%%%%%%%%%%%%%%%%%%%%%%%%%%%%%%%%%%%%%%%%%%%%%%%%%%%%%%%%%%%%%%%%%%%%%%%%%%%%%%%%%%%%%%%%%%%%%%%%%%%%%%%%%%%%%%%%%%%%%%%%%%%%%%%%%%%%%%%%%%%%%%%%%%%%%%%%%%%%%%

\section{Density operator}

\setcounter{de}{0}
\setcounter{tw}{0}
\setcounter{lem}{0}
\setcounter{co}{0}

Let us consider a quantum system  on a separable Hilbert space ${\cal H}.$  
Assume that $\{|\varphi_j \big>\}_{j=1}^{\dim {\cal H}}$ is a complete set of orthonormal vectors in the Hilbert space ${\cal H}.$ We know only a
probability of detecting the system in each of the states $|\varphi_j\big>.$ To represent a state of the system we introduce a density operator.
Following von Neumann \cite{neu} 
\begin{de}
\label{def1}
The operator given by
\[
\hat{\varrho}:= {\rm u-}\, \lim_{n \rightarrow \dim {\cal H}} \sum_{j=1}^n p_j  |\varphi_j\big> \big<\varphi_j|,\;\; 
\forall \, j \;\; p_j \geq 0\;\;, \;\; \sum_{j=1}^{\dim {\cal H}} p_j=1
\]
is called a {\bf density operator}. Each number $p_j, \, j=1,2,\ldots,\dim {\cal H} $ is the probability of observing the system in the state
represented by a ket $|\varphi_j\big>.$ If one of these numbers equals $1$ we say that the system is in a {\bf pure state}. Otherwise the system is in
a {\bf mixed state}.
\end{de}
The symbol ${\rm u-}$ denotes the uniform convergence of a sequence of operators.

Definition \ref{def1} presents a  method of construction of a density operator $\hat{\varrho}.$ 
Its equivalent formulation,  in spirit of considerations  presented in the Introduction, is the following.
\begin{de}
\label{def2}
An operator $\hat{\varrho}: {\cal H} \rightarrow {\cal H}$  is a {\bf density operator} if it is:
\begin{enumerate}
\item
  positive i.e. $\forall \, |\phi \big> \,\in \,{\cal H} \;\;  \big<\phi|\hat{\varrho}|\phi \big> \,\geq 0, $
\item
self -- adjoint $\hat{\varrho}^+=\hat{\varrho}$ and
\item
its trace ${\rm Tr} \hat{\varrho}=1.$
\end{enumerate}
\end{de}
As we deal with operators acting in general in an infinite dimensional space, it seems to be necessary to comment on a notion of trace. For any bounded
and positive operator $\hat{A}$ its trace is defined as \cite{ree}
\[
{\rm Tr} \hat{A}:= \sum_{i=1}^{\dim {\cal H}} \big< \varphi_i| \hat{A}|\varphi_i \big>
\] 
and is independent of the orthonormal basis chosen. So defined functional has properties analogous to the trace of a finite dimensional $n \times n$
matrix. Hence
\[
{\rm Tr}(\hat{A}+ \hat{B})= {\rm Tr}\hat{A} + {\rm Tr}\hat{B} \;\; , \;\;
\forall \, b\geq 0 \;\;{\rm Tr}(b\hat{A})= b \cdot {\rm Tr} \hat{A}.
\]
Moreover, the trace is invariant under any unitary transformation
\[
{\rm Tr} (\hat{U}\hat{A}\hat{U}^{-1})= {\rm Tr}\hat{A}.
\]

From Defs. \ref{def1} and \ref{def2} we can deduce several properties of the density operator. First of all we will see that the operator $\hat{\varrho}$ is
bounded. Indeed, by definition its domain is the whole Hilbert space ${\cal H}.$
Moreover,  it is self -- adjoint. Hence, from the Hellinger -- Toeplitz theorem \cite{ree} we obtain that the density operator is bounded.

As it can be seen from Def. \ref{def1} and from the spectral theorem,  
eigenvalues of the density operator are nonnegative and they do not exceed $1$.
Since the norm $||\hat{\varrho}||$ is the supremum of the eigenvalues of the operator $\hat{\varrho},$ it satisfies the inequality $||\hat{\varrho}||
\leq 1.$

From Def. \ref{def1} we easily calculate that ${\rm Tr} \hat{\varrho}^2 \leq 1.$ Thus the density operator is a Hilbert -- Schmidt operator. Its
Hilbert -- Schmidt norm, given by
\[
|| \hat{\varrho} ||_2:=\sqrt{{\rm Tr (\hat{\varrho}^+ \hat{\varrho} )}},
\]
is not greater than $1.$ Moreover, as it is well known, $|| \hat{\varrho} ||_2 = 1$ if and only if the density operator represents a pure state. The
Hilbert -- Schmidt type operators constitute a $*\;$-- ideal in the algebra $B({\cal H}).$

Finally, as the eigenvalues of the density operator $\hat{\varrho}$ are nonnegative, the density operator is a trace class operator and its trace norm
\[
|| \hat{\varrho} ||_1:= {\rm Tr} \sqrt{\hat{\varrho}^* \hat{\varrho}}= {\rm Tr}|\hat{\varrho}|= {\rm Tr}\hat{\varrho}=1.
\]

The space of trace class operators $B_1({\cal H})$ is also a $*\;$-- ideal in the algebra $B({\cal H}).$ For each trace class operator $\hat{A}$ the following
estimation holds
\be
\label{n1}
||\hat{A}|| \leq ||\hat{A}||_2 \leq ||\hat{A}||_1.
\ee
Since every density operator is a trace class operator, it is compact. For two arbitrary operators $\hat{A} \in B_1(\cal H)$ and $\hat{B} \in B(\cal
H)$ the trace of the product is abelian
\[
{\rm Tr} (\hat{A} \hat{B})= {\rm Tr} (\hat{B} \hat{A}).
\]
The density operator is positive. Hence
for every operator $\hat{A} \in B({\cal H})$  the mean value of the product obeys 
\[
\big< \hat{A}\hat{A}^+ \big> = {\rm Tr} (\hat{\varrho} \hat{A} \hat{A}^+)\geq 0.
\]
In our paper we consider finite and infinite dimensional separable Hilbert spaces. In that second case two realisations: the set $l^2$ of one -- column
complex matrices and the space $L^2$ of square integrable functions over ${\mathbb R}$ are analysed. We discuss mixed and pure states.
\vspace{0.5cm}

Assume that a matrix representation $ \Big[ \big<\varphi_i | \hat{\varrho}| \varphi_j \big> \Big]_1^{\dim {\cal H}}$ of an operator $\hat{\varrho}$ is known. The kets $\{| \varphi_i
\big>\}_{i=1}^{\dim {\cal H}}$ constitute an orthonormal basis of the Hilbert space ${\cal H}.$
We intend to settle whether this matrix represents a physical state  of a quantum system.

Let us consider a finite dimensional case $\dim {\cal H} < \infty $ first. Then the matrix $\Big[\big<\varphi_i | \hat{\varrho}| \varphi_j \big>\Big]_1^{\dim {\cal H}}$ completely
determines the operator $\hat{\varrho},$ which is defined on the whole space ${\cal H}$ and is bounded. Therefore we use Def. \ref{def2}. 
At the beginning  we test if
the matrix is symmetric \\
$\forall_{i,j}\;\;\; \big<\varphi_i | \hat{\varrho}| \varphi_j \big>= \overline{\big<\varphi_j | \hat{\varrho}| \varphi_i \big>}$ and its trace $\sum_{i=1}^{\dim {\cal H}}
\big<\varphi_i | \hat{\varrho}| \varphi_i \big>=1.$ Then calculating the principal minors we check whether the matrix is positive. When the matrix $\Big[\big<\varphi_i | \hat{\varrho}| \varphi_j \big> \Big]_1^{\dim {\cal H}}$ passes these three tests we conclude that it really represents a density operator. 

For a pure state this procedure becomes simpler. A matrix $\Big[\big<\varphi_i | \hat{\varrho}| \varphi_j \big>\Big]_1^{\dim {\cal H}}$ represents a pure state if it is symmetric,
its trace equals $1$ and the square of it is the same matrix
\[
\forall_{i,j} \;\;\;
\sum_{k=1}^{\dim{\cal H}}\big<\varphi_i|\hat{\varrho}|\varphi_k\big>\big<\varphi_k|\hat{\varrho}|\varphi_j\big>=
\big<\varphi_i|\hat{\varrho}|\varphi_j\big>.
\]

The case of a separable but infinite dimensional Hilbert space is more elaborated. It may happen that in a given orthonormal basis
$\{| \varphi_i \big>\}_{i=1}^{\infty}$ the matrix $\Big[\big<\varphi_i | \hat{A}| \varphi_j \big>\Big]_1^{ \infty}$ of an operator $\hat{A}$ exists but this matrix does not uniquely
characterise this operator (compare \cite{ach}).

Hence to know whether a matrix $\Big[\big<\varphi_i|\hat{\varrho}|\varphi_j\big>\Big]_1^{ \infty}$ can represent a density operator, we propose to check first if the operator
$\hat{\varrho}$ is a Hilbert -- Schmidt operator. As it was proved in \cite{ach}, if
\[
\sum_{i,j=1}^{\infty}|\big<\varphi_i|\hat{A}|\varphi_j\big>|^2< \infty
\]
then the matrix $[\big<\varphi_i|\hat{A}|\varphi_j\big>]$ represents an operator of the Hilbert -- Schmidt type. Any Hilbert -- Schmidt operator
$\hat{A}$ is bounded and  is defined on the whole space ${\cal H}.$ Moreover, in this case the matrix $\Big[\big<\varphi_i|\hat{A}|\varphi_j\big>\Big]_1^{ \infty}$
completely characterises the operator $\hat{A}.$ In addition, if the matrix $\Big[\big<\varphi_i|\hat{A}|\varphi_j\big>\Big]_1^{ \infty}$ is symmetric, we conclude that the
operator $\hat{A}$ is self -- adjoint.

Suppose that  a matrix  $\Big[\big<\varphi_i|\hat{\varrho}|\varphi_j\big>\Big]_1^{ \infty}$ is symmetric and of the Hilbert -- Schmidt type.
In the next step we have to test if the matrix  $\Big[\big<\varphi_i|\hat{\varrho}|\varphi_j\big>\Big]_1^{ \infty}$ represents a positive operator. 
We start from a few lemmas.
\begin{lem}
If a Hilbert -- Schmidt operator $\hat{A}$ is positive, then in any matrix representation $\Big[\big<\varphi_i|\hat{A}|\varphi_j\big>\Big]_1^{ \infty}$ all diagonal elements
$\big<\varphi_i|\hat{A}|\varphi_i\big>$ are nonnegative.
\end{lem}
Any Hilbert -- Schmidt operator $\hat{A}$ is bounded.
Every bounded positive operator is self -- adjoint \cite{ree}. Moreover, each of its matrix representations determines this operator. Hence all the diagonal
elements $\big<\varphi_i|\hat{A}|\varphi_i\big>$ are real. From the definition of a positive operator for every $i \in {\cal N}$ there is
\[
 \big<\varphi_i|\hat{A}|\varphi_i\big> \geq  0.  \;\;\; \rule{2mm}{2mm}
\]
Therefore we conclude that  
\begin{co}
A necessary condition for a Hilbert -- Schmidt operator $\hat{\varrho}$ to be a density operator is that in any matrix representation
$\Big[\big<\varphi_i|\hat{\varrho}|\varphi_j\big>\Big]_1^{ \infty}$ all diagonal elements $\big<\varphi_i|\hat{\varrho}|\varphi_i\big>$  are nonnegative.
\end{co}  
Another observation is the following.
\begin{lem}
If $\hat{\varrho}$ is a density operator then $||\hat{\varrho}- \hat{1}|| \leq 1.$
\end{lem}
This statement is the straightforward consequence of the observation that the eigenvalues of the operator $\hat{\varrho}- \hat{1}$ belong to the
interval $[-1,0].$ As the norm of an arbitrary operator $\hat{A}$ is the supremum of the absolute value of its eigenvalues, the Lemma
is proved. \rule{2mm}{2mm}

\begin{lem}
\label{lem3}
If $\hat{\varrho}$ is a density operator then the uniform limit of the sequence $\{( \hat{1}-\hat{\varrho})^{n}\}_{n=1}^{\infty}$ of operators is the
projective operator on the kernel of the density operator.
\end{lem}
By Def. \ref{def1} eigenvalues of the operator  $(\hat{1}-\hat{\varrho})$ are nonnegative and they
  do not exceed $1.$ Thus in the uniform
limit we obtain a sum of $1$-dimensional mutually orthogonal projective operators. This sum represents an orthogonal projection on the kernel of
$\hat{\varrho}.$
 \rule{2mm}{2mm}

The straightforward consequences of this lemma are two observations.
\begin{co}
If $\hat{\varrho}$ is a density operator then
\[
\Big({\rm u}- \lim_{n \rightarrow \infty}  (\hat{1}- \hat{\varrho})^n \Big)^2
= 
{\rm u}- \lim_{n \rightarrow \infty} (\hat{1}- \hat{\varrho})^n .
\]
\end{co}
\begin{co}
For every density operator  $\hat{\varrho}$
\[
{\rm Tr}\Big( ({\rm u}- \lim_{n \rightarrow \infty} (\hat{1}- \hat{\varrho})^n) \cdot \hat{\varrho}\Big)=0.
\]
\end{co}

Therefore we propose the following criterion for a matrix $ \Big[\big<\varphi_i|\hat{\varrho}|\varphi_j\big>\Big]_1^{ \infty}$ to represent a density operator
$\hat{\varrho}$.
\begin{tw}
\label{tw3.0}
A matrix   $\Big[ \big<\varphi_i|\hat{\varrho}|\varphi_j\big>\Big]_1^{ \infty}$ represents a quantum state iff:
\begin{enumerate}
\item
$
\sum_{i,j=1}^{\infty}
|\big<\varphi_i|\hat{\varrho}|\varphi_j\big>|^2 \leq 1,$
\item
$
\forall \; 1 \leq i,j < \infty \;\; \big<\varphi_i|\hat{\varrho}|\varphi_j\big>= \overline{\big<\varphi_j|\hat{\varrho}|\varphi_i\big>}$,
\item
the sequence  $\Big\{ \Big[ \big<\varphi_i| (\hat{1} - \hat{\varrho})^n |\varphi_j\big>\Big]_1^{ \infty} \Big\}_{n=1}^{\infty} $ is convergent in the norm,
\item
$
\sum_{i=1}^{\infty}
\big<\varphi_i|\hat{\varrho}|\varphi_i\big> = 1.$
\end{enumerate}
\end{tw}
\noindent
`$\Rightarrow$'
\newline
As it was shown, a density operator is a Hilbert -- Schmidt operator and the trace of its square is not greater than $1.$ Moreover, it is self -- adjoint.
Lemma \ref{lem3} implies the  property $(iii)$. By definition the trace of the density operator equals $1.$
\newline
`$\Leftarrow$'
\newline
The first condition says that the operator $\hat{\varrho}$ is of a Hilbert -- Schmidt type. As it is also symmetric, we see that it is self -- adjoint.
Thus from the spectral theorem it can be represented by a diagonal matrix and from $(i)$ absolute values of its eigenvalues do not exceed $1.$
Therefore the matrix $\Big[\big<\varphi_i| \hat{1}- \hat{\varrho}|\varphi_j\big>\Big]_1^{ \infty}$ is also diagonal and the numbers on the diagonal belong to the interval
$[0,2].$ But if one or more elements of the diagonal matrix $\Big[\big<\varphi_i| \hat{1}- \hat{\varrho}|\varphi_j\big>\Big]_1^{ \infty}$ are greater than $1,$ the sequence
of matrices $\Big\{\;\Big [\big<\varphi_i|(\hat{1}- \hat{\varrho})^n|\varphi_j\big> \Big]_1^{ \infty} \;\Big\}_{n=1}^{\infty}$ is divergent. Hence the convergence of this sequence
implies that all of eigenvalues of the operator $\hat{\varrho}$ are nonnegative and the operator is positive. Thus its trace is well defined and if the
condition $(iv)$ holds, we see that $\hat{\varrho}$ is really a density operator. \rule{2mm}{2mm}

Another criterion is based on a notion of  square root of an operator.

\begin{lem}{\cite{ree}}
\label{tw2.4}
Let $\hat{A}$ be a positive and self -- adjoint linear operator defined on the Hilbert space ${\cal H }.$ Moreover, let $|| \hat{A}|| \leq 1.$ Then there
exists a unique positive and self -- adjoint linear operator ${\hat B}$ such that $\hat{B}^2= \hat{A}.$ The operator ${\hat B}$ is a uniform limit of the
series
\[
\hat{B}= \sqrt{\hat{A}}= \sqrt{\hat{1}+ (\hat{A}-\hat{1})}= \hat{1} + \frac{1}{2}(\hat{A}-\hat{1})
- \frac{1}{8}(\hat{A}-\hat{1})^2 + \frac{1}{16}(\hat{A}-\hat{1})^3 + \ldots
\]
\be
\label{2.4}
 + \frac{(-1)^{n+1}(2n-3)!!}{n! \, 2^n} (\hat{A}-\hat{1})^n + \ldots 
\ee 
\end{lem}
A proof of this statement can be found in \cite{ree}. The series is convergent only for operators satisfying the estimation $||\hat{A}||\leq 1.$  
Thus
\begin{tw}
\label{tw4}
A matrix   $ \Big[\big<\varphi_i|\hat{\varrho}|\varphi_j\big> \Big]_1^{ \infty}$ represents a quantum state iff:
\begin{enumerate}
\item
$
\sum_{i,j=1}^{\infty}
|\big<\varphi_i|\hat{\varrho}|\varphi_j\big>|^2 \leq 1,$
\item
$
\forall \; 1 \leq i,j < \infty \;\; \big<\varphi_i|\hat{\varrho}|\varphi_j\big>= \overline{\big<\varphi_j|\hat{\varrho}|\varphi_i\big>}$,
\item
the series
\[
\Big[ \big<\varphi_i|\hat{1}|\varphi_j\big> \Big]_1^{ \infty} + \frac{1}{2} \Big[\big<\varphi_i|\hat{\varrho}-\hat{1}|\varphi_j\big>\Big]_1^{ \infty}
- \frac{1}{8} \Big[\big<\varphi_i|(\hat{\varrho}-\hat{1})^2|\varphi_j\big>\Big]_1^{ \infty} + \frac{1}{16} \Big[\big<\varphi_i|(\hat{\varrho}-\hat{1})^3|\varphi_j\big> \Big]_1^{ \infty} + \ldots
\]
\be
\label{tw4w1}
 + \frac{(-1)^{n+1}(2n-3)!!}{n! \, 2^n} \Big[\big<\varphi_i|(\hat{\varrho}-\hat{1})^n|\varphi_j\big>\Big]_1^{ \infty} + \ldots 
\ee 
 is convergent in the norm to $\sqrt{\Big[  \big<\varphi_i|\hat{\varrho}|\varphi_j\big>   \Big]_1^{ \infty} } $,
\item
$
\sum_{i=1}^{\infty}
\big<\varphi_i|\hat{\varrho}|\varphi_i\big> = 1.$
\end{enumerate}
\end{tw}

A proof of this theorem is analogous to the proof of Theorem \ref{tw3.0}. The only  difference refers to the series defining the square root of the
matrix $ \Big[ \big<\varphi_i|\hat{\varrho}|\varphi_j\big> \Big]_1^{ \infty}.$ If the operator $\hat{\varrho}$ is indeed a density operator, then it is positive and 
$||\hat{\varrho}|| \leq 1.$ Thus its square root can be calculated as the sum of the series (\ref{tw4w1}).

On the other hand, the series (\ref{tw4w1}) defines the square root of an operator with the norm not greater than $1.$ Therefore if this series is
convergent, the operator $\hat{\varrho}$ is positive.

By a slight modification of formula (\ref{2.4}) we obtain that the square root of a positive self -- adjoint operator $\hat{A}$ satisfying
$||\hat{A}||\leq 1$ equals
\be
\label{2.5}
\sqrt{\hat{A}}=
\frac{1}{2} \hat{A}+ \sum_{l=2}^{\infty}\frac{(-1)^{l+1}}{l!} \frac{(2l-3)!!}{2^l} \sum_{r=0}^{l-1}(-1)^r \left( \begin{array}{c} l \\ r
\end{array}\right) \hat{A}^{l-r}.
\ee 
If the operator $\hat{A}$ is of a trace class  so is each element of the series  (\ref{2.5}).

\begin{tw}
\label{tw4.0}
A matrix   $\Big[ \big<\varphi_i|\hat{\varrho}|\varphi_j\big>\Big]_1^{ \infty}$ represents a quantum state iff:
\begin{enumerate}
\item
$
\sum_{i,j=1}^{\infty}
|\big<\varphi_i|\hat{\varrho}|\varphi_j\big>|^2 \leq 1,$
\item
$
\forall \; 1 \leq i,j < \infty \;\; \big<\varphi_i|\hat{\varrho}|\varphi_j\big>= \overline{\big<\varphi_j|\hat{\varrho}|\varphi_i\big>}$,
\item
the series 
\be
\label{5}
\frac{1}{2}\Big[\big<\varphi_i|\hat{\varrho}^2|\varphi_j\big> \Big]_1^{ \infty}+ \sum_{l=2}^{\infty}\frac{(-1)^{l+1}}{l!} \frac{(2l-3)!!}{2^l} \sum_{r=0}^{l-1}(-1)^r \left(
\begin{array}{c} l \\ r \end{array}\right) \Big[ \big<\varphi_i|\hat{\varrho}^{2(l-r)}|\varphi_j\big>\Big]_1^{ \infty}
\ee
 is convergent to the operator $\hat{\varrho}$ in the trace norm,
\item
$
\sum_{i=1}^{\infty}
\big<\varphi_i|\hat{\varrho}|\varphi_i\big> = 1.$
\end{enumerate}
\end{tw}
The  condition $(iii)$ says that $\hat{\varrho}= \sqrt{\hat{\varrho}^2}.$ In fact it is sufficient to check this relationship in the sense of the uniform
convergence of the series (\ref{5}). However, this is a quite complex procedure. In this case it is much easier  to test the trace convergence.
\newline
`$\Leftarrow$'
\newline
 Let us introduce a new symbol
\[
\hat{\varrho}_n:= \frac{1}{2} \hat{\varrho}^2+ \sum_{l=2}^{n}\frac{(-1)^{l+1}}{l!} \frac{(2l-3)!!}{2^l} \sum_{r=0}^{l-1}(-1)^r \left( \begin{array}{c}
l \\ r \end{array}\right) \hat{\varrho}^{2(l-r)}.
\]
When the sequence $\{( \hat{\varrho}_n - \hat{\varrho})\}_{n=1}^{\infty}$ is trace convergent then from the estimation (\ref{n1}) it is also uniform convergent. Thus the operator $\hat{\varrho}$ is positive.
\newline
`$\Rightarrow$'
\newline
On the other hand we can consider a diagonal form of the matrix  $ \Big[\big<\varphi_i |\hat{\varrho} |\varphi_j \big>\Big]_1^{ \infty}$
representing a density operator.
For each index $ i $ the sequence $\{\big<\varphi_i | \hat{\varrho}_n|\varphi_i\big>\}_{n=1}^{\infty}$ is convergent to $\big<\varphi_i |
\hat{\varrho}|\varphi_i \big>.$ Moreover, each number $\big<\varphi_i | \hat{\varrho}_n|\varphi_i\big>$ is nonnegative and the sequence
$\{\big<\varphi_i | \hat{\varrho}_n|\varphi_i\big>\}_{n=1}^{\infty}$ is growing for every $i.$ Thus for every $n$ the operator $(\hat{\varrho}- \hat{\varrho}_n)$ is positive and its trace norm is simply the trace of  the operator  $(\hat{\varrho}- \hat{\varrho}_n).$ 
The sequence of numbers $\{||\hat{\varrho}_n-\hat{\varrho}||_1 \}_{n=1}^{\infty}$ tends to $0$ so the uniform convergence of the sequence
$\{\hat{\varrho}_n\}_{n=1}^{\infty}$ implies also its trace convergence to $\hat{\varrho}.$ \rule{2mm}{2mm}

The previous theorem leads to, perhaps, the most useful criterion formulated below.
\begin{tw}
\label{tw5.0}
A matrix   $ \Big[\big<\varphi_i|\hat{\varrho}|\varphi_j\big> \Big]_1^{ \infty}$ represents a quantum state iff:
\begin{enumerate}
\item
$
\sum_{i,j=1}^{\infty}
|\big<\varphi_i|\hat{\varrho}|\varphi_j\big>|^2 \leq 1,$
\item
$
\forall \; 1 \leq i,j < \infty \;\; \big<\varphi_i|\hat{\varrho}|\varphi_j\big>= \overline{\big<\varphi_j|\hat{\varrho}|\varphi_i\big>}$,
\item
the series 
\be
\label{6}
\frac{1}{2} \sum_{i=1}^{\infty} \big<\varphi_i|\hat{\varrho}^2|\varphi_i\big>+ \sum_{l=2}^{\infty}\frac{(-1)^{l+1}}{l!} \frac{(2l-3)!!}{2^l}
\sum_{r=0}^{l-1}(-1)^r \left( \begin{array}{c} l \\ r \end{array}\right) \sum_{i=1}^{\infty} \big<\varphi_i|\hat{\varrho}^{2(l-r)}|\varphi_i\big> =1,
\ee
\item
$
\sum_{i=1}^{\infty}
\big<\varphi_i|\hat{\varrho}|\varphi_i\big> = 1.$
\end{enumerate}
\end{tw}
\noindent
`$\Rightarrow$'
\newline
A density operator $\hat{\varrho}$ is of a Hilbert -- Schmidt type and self -- adjoint. Moreover, it is positive, so
$\sqrt{\hat{\varrho}^2}=\hat{\varrho}.$ As $||\hat{\varrho}|| \leq 1,$ the square root of the operator $\hat{\varrho}$ is determined by the series
(\ref{2.4}). Thus ${\rm Tr}\sqrt{\hat{\varrho}^2}=1.$
\newline
`$\Leftarrow$'
\newline
From the conditions $(i)$ and $(ii)$ we obtain that an operator $\hat{\varrho}$ is of a Hilbert -- Schmidt type and is self -- adjoint. Its eigenvalues do
not exceed $1.$ The square of a Hilbert -- Schmidt operator is a trace class operator. Moreover, the set of trace class operators is an ideal in the
space $B({\cal H}).$ Thus the trace of each operator $\hat{\varrho}^{2n}$ is well defined. The sum (\ref{6})  equals ${\rm Tr}\sqrt{\hat{\varrho}^2}.
$ If ${\rm Tr}\sqrt{\hat{\varrho}^2}= {\rm Tr}\hat{\varrho},$ there must be $\sqrt{\hat{\varrho}^2}=\hat{\varrho}.$ Hence $\hat{\varrho}$ is a positive
operator. As its trace equals $1,$ it is a density operator.
\rule{2mm}{2mm}

The last two criteria presented in this section are based on a clever idea proposed by M. Wasiak \cite{mw}.
\begin{tw}
\label{tw4.1}
A matrix   $ \Big[ \big<\varphi_i|\hat{\varrho}|\varphi_j\big>\Big]_1^{ \infty}$ represents a quantum state iff:
\begin{enumerate}
\item
$
\sum_{i,j=1}^{\infty}
|\big<\varphi_i|\hat{\varrho}|\varphi_j\big>|^2 \leq 1,$
\item
$
\forall \; 1 \leq i,j < \infty \;\; \big<\varphi_i|\hat{\varrho}|\varphi_j\big>= \overline{\big<\varphi_j|\hat{\varrho}|\varphi_i\big>}$,
\item
for every natural number $n$ the sum 
\[
\sum_{k=0}^n (-1)^k \left( \begin{array}{c} n \\ k \end{array} \right) \sum_{i=1}^{\infty}   \big<\varphi_i|\hat{\varrho}^{k+1}|\varphi_i\big> \geq 0,
\]
\item
$
\sum_{i=1}^{\infty}
\big<\varphi_i|\hat{\varrho}|\varphi_i\big> = 1.$
\end{enumerate}
\end{tw}
The conditions $(i)$, $(ii)$ and $(iv)$ were considered before. Assuming that they are fulfilled, we can choose a basis $\{|\varphi_i \big>\}^{\infty}_{i=1},$ in
which the matrix $ \Big[\big<\varphi_i| \hat{\varrho}| \varphi_j \big> \Big]_1^{ \infty}$ is diagonal. Nonnegative eigenvalues of the operator $\hat{\varrho}$ are denoted by
$x_i,$ negative by $-y_i.$

Since $
\sum_{i=1}^{\infty}
|\big<\varphi_i|\hat{\varrho}|\varphi_i\big>|^2 \leq 1,$ we see that all $ x_i \leq 1$ and all $y_i \leq 1.$ 
\[
\sum_{k=0}^n (-1)^k \left( \begin{array}{c} n \\ k \end{array} \right) {\rm Tr} \hat{\varrho}^{k+1}=
\sum_{k=0}^n (-1)^k \left( \begin{array}{c} n \\ k \end{array} \right) \sum_{i=1}^{\infty}\big( x_i \cdot x_i^k- y_i\cdot (-1)^k y_i^k \big)=
\]
\[
= \sum_{i=1}^{\infty} x_i \sum_{k=0}^n \left( \begin{array}{c} n \\ k \end{array} \right) (-x_i)^k-
\sum_{i=1}^{\infty} y_i \sum_{k=0}^n \left( \begin{array}{c} n \\ k \end{array} \right) (y_i)^k=
\]
\[
= \sum_{i=1}^{\infty} x_i(1-x_i)^n - \sum_{i=1}^{\infty} y_i(1+y_i)^n. 
\]
As $n$ tends to $\infty,$ the first component of the sum goes to $0$. On the contrary, the sum $\sum_{i=1}^{\infty} y_i(1+y_i)^n$ grows for $n
\rightarrow \infty.$ \rule{2mm}{2mm}

An obvious consequence of Theorem \ref{tw4.1} and its proof is the following statement.
\begin{tw}
\label{co4.1}
A matrix   $ \Big[\big<\varphi_i|\hat{\varrho}|\varphi_j\big> \Big]_1^{ \infty}$ represents a quantum state iff:
\begin{enumerate}
\item
$
\sum_{i,j=1}^{\infty}
|\big<\varphi_i|\hat{\varrho}|\varphi_j\big>|^2 \leq 1,$
\item
$
\forall \; 1 \leq i,j < \infty \;\; \big<\varphi_i|\hat{\varrho}|\varphi_j\big>= \overline{\big<\varphi_j|\hat{\varrho}|\varphi_i\big>}$,
\item
the limit 
\[
\lim_{n \rightarrow \infty} \; \sum_{k=0}^n (-1)^k \left( \begin{array}{c} n \\ k \end{array} \right) \sum_{i=1}^{\infty}
\big<\varphi_i|\hat{\varrho}^{k+1}|\varphi_i\big> =0,
\]
\item
$
\sum_{i=1}^{\infty}
\big<\varphi_i|\hat{\varrho}|\varphi_i\big> = 1.$
\end{enumerate}
\end{tw}
Presented criteria are of course valid not only for mixed but also for pure states. However, in this latter case the simplest test seems to be the
following procedure.
\begin{tw}
\label{tw4.4}
A matrix   $ \Big[ \big<\varphi_i|\hat{\varrho}|\varphi_j\big> \Big]_1^{ \infty}$ represents a pure quantum state iff:
\begin{enumerate}
\item
$
\sum_{i,j=1}^{\infty}
|\big<\varphi_i|\hat{\varrho}|\varphi_j\big>|^2 = 1,$
\item
$
\forall \; 1 \leq i,j < \infty \;\; \big<\varphi_i|\hat{\varrho}|\varphi_j\big>= \overline{\big<\varphi_j|\hat{\varrho}|\varphi_i\big>}$,
\item
$
\sum_{j=1}^{\infty} \big<\varphi_i|\hat{\varrho}|\varphi_j\big> \big<\varphi_j|\hat{\varrho}|\varphi_k\big>
=\big<\varphi_i|\hat{\varrho}|\varphi_k\big>,
$
\item
$
\sum_{i=1}^{\infty}
\big<\varphi_i|\hat{\varrho}|\varphi_i\big> = 1.$
\end{enumerate}
\end{tw}
We begin with checking if the operator $\hat{\varrho}$ is a Hilbert -- Schmidt type, because only in this case the matrix $
\big<\varphi_i|\hat{\varrho}|\varphi_j\big>$ completely characterises the operator. Moreover, as we already mentioned, a symmetric Hilbert -- Schmidt
operator is self -- adjoint. In the  step $(iii)$ we test, whether the operator $\hat{\varrho}$ is a projection.

\vspace{0.5cm}

All of our previous considerations done for infinite dimensional matrices can be adapted for the Hilbert space $L^2$ of square integrable functions
over ${\mathbb R}$. Indeed, if functions
\[
f(x), g(x) \in L^2\;\; {\rm and} \;\;
g(y)= \hat{A} f(x),
\]
 we write
\[
g(y)= \int_{\mathbb R}A(y,x) f(x) dx.
\]
Thus an operator $\hat{A}$ is symmetric if
\[
 \forall\; x,y \in {\mathbb R}\;\; A(x,y)=\overline{A(y,x)}.
\]
 An operator $\hat{A}$ is a Hilbert-- Schmidt operator if 
\[
\int_{{\mathbb R}^2} |A(x,y)|^2dxdy < \infty.
\]
A product of operators is calculated as
\[
\big(A \cdot B\big) (x,z)=\int_{\mathbb R}A(x,y)B(y,z)dy
\]
and a trace of an operator $\hat{A}$ can be found from the relation
\[
{\rm Tr} \hat{A}= \int_{\mathbb R}A(x,x)dx.
\]

%%%%%%%%%%%%%%%%%%%%%%%%%%%%%%%%%%%%%%%%%%%%%%%%%%%%%%%%%%%%%
%%%%%%%%%%%%%%%%%%%%%%%%%%%%%%%%%%%%%%%%%%%%%%%%%%%%%%%%%%%%%%%%%%%%%%%%%%%%%%%%%%%%
\section{Wigner function}
When the phase space formulation of quantum mechanics is considered,  two fundamental elements: a phase space and a $*$-- product must be taken into account.  Examples  of constructing a phase space can be found in \cite{varilly1} -- \cite{ja2}. In this section we restrict ourselves to problems, in  which phase spaces are  differentiable symplectic manifolds.

There exists  vast literature devoted to existence and construction of $*$-- products. A systematic treatment can be found in  \cite{dit} -- \cite{bor}.
On every symplectic manifold $({\cal M}, \omega)$ there exists a nontrivial $*$-- product \cite{wild} --  \cite{omor}. An iterative construction of the natural $*$-- product of the Weyl type has been proposed by Fedosov \cite{6}, \cite{7}. We assume that a $*$-- product is local and in its differential form is of the Weyl type. However, we see the necessity of applying an integral form of the $*$-- product too. This  postulate will become clear when we discuss some properties of Wigner eigenfunctions.

An observable on a phase space $({\cal M}, \omega)$ is any smooth real  function on ${\cal M}$ being a formal series in the Dirac constant $\hbar$
\be
\label{10.5}
C^{\infty}({\cal M})[[\hbar]] \ni A(q^1, \ldots, q^{2n})= \sum_{i=0}^{\infty} \hbar^i A_i(q^1, \ldots, q^{2n}).
\ee
By $q^1, \ldots, q^{2n}$ we denote local coordinates on the symplectic manifold  $({\cal M}, \omega).$ Each function $A_i(q^1, \ldots, q^{2n}), \; \dim{\cal M}=2n,$ is a  smooth real function on $({\cal M}, \omega).$ 

Following the scheme presented in the Introduction, we  choose a $*$-- algebra ${\cal A}.$ Our choice ensures that 
 all smooth functions on $({\cal M}, \omega),$ which are formal series in $\hbar$ and have  compact supports, belong to ${\cal A}.$ 
We will return to the question of choice and structure of this algebra later on when a Stratonovich -- Weyl correspondence will be discussed.

The involution
`$*$' is realised by the complex conjugation. A product in the algebra ${\cal A}$ is a Weyl type $*$-- product.

According to the general definition quoted in the Introduction, quantum states are positive linear functionals over the algebra  ${\cal A}$ satisfying the normalization condition (\ref{10}). As it was shown in \cite{7}, every such a  functional $f$ can be written in the following form
\[
\forall \; A(q^1, \ldots, q^{2n}) \in {\cal A}
\]
\[
f\Big(A(q^1, \ldots, q^{2n})\Big) =
\int_{\cal M}A(q^1, \ldots, q^{2n})*W(q^1, \ldots, q^{2n}) t(q^1, \ldots, q^{2n})\omega^n=
\]
\be
\label{11}
=\int_{\cal M}W(q^1, \ldots, q^{2n})*A(q^1, \ldots, q^{2n}) t(q^1, \ldots, q^{2n})\omega^n.
\ee
A real function
\[
C^{\infty}({\cal M})[[\hbar]] \ni
t(q^1, \ldots, q^{2n})=\sum_{i=0}^{\infty} \hbar^i t_i(q^1, \ldots, q^{2n} )
\]
is called a {\bf trace density}. An explicit construction of this function  was proposed by Fedosov \cite{9}. The trace density is  determined by a symplectic curvature tensor and its derivatives. A function $W(q^1, \ldots, q^{2n})$ contains information about the state. This function is called a {\bf Wigner function}. 
The integral
\be
\label{12}
\frac{1}{(2 \pi \hbar)^n}\int_{\cal M}A(q^1, \ldots, q^{2n}) t(q^1, \ldots, q^{2n})\omega^n
\ee
is often called a {\bf trace} as  is a classical counterpart of the trace of  operator. 

A functional nowadays called a Wigner function on the phase space ${\mathbb R}^{2n}$ appeared in the literature eighty years ago  \cite{wig}.
Since in the case of a system with the phase space ${\mathbb R}^{2n}$ an explicit form of a mapping between operators and functions is known \cite{ja1}, several properties of a Wigner function can be directly obtained from a density operator. They have been published in  \cite{tat} -- \cite{lee}, \cite{tak} and \cite{dias4}. But even for systems modelled on   the phase space ${\mathbb R}^{2n}$ the question of admissible states was considered only in  \cite{dias1}. Criteria of identification of pure states were proposed in \cite{dias2} and \cite{ja3}.

In a general situation we do not know an explicit form of a mapping between operators acting in a Hilbert space and functions in a phase space. Yet, as both: the Hilbert space  and the phase space formulation describe the same physical reality, there must be a correspondence between them. This correspondence implies an algebra isomorphism between the quantum algebra of operators ${\cal A}_{\cal H}$ and the algebra of functions  ${\cal A}$ on the phase space. This isomorphism is known as a Stratonovich -- Weyl correspondence \\ $SW: \;{\cal A}_{\cal H} \rightarrow {\cal A},$ \cite{stra}. Below we list  properties of the Stratonovich -- Weyl correspondence.
\renewcommand{\theenumi}{ \arabic{enumi}}
\begin{enumerate}
\item
The  mapping 
 is one -- to -- one.
\newline
\noindent
We stress a relationship between the algebras ${\cal A}_{\cal H}$  and ${\cal A}.$ The choice of the algebra of operators ${\cal A}_{\cal H}$ determines the choice of the algebra of functions ${\cal A}.$ On the other hand we know, that this choice is not unique and for different algebras  ${\cal A}_{\cal H}$ or equivalently ${\cal A}$ we obtain the same quantum mechanics.
\item
The  $SW$ correspondence is linear. Moreover, $SW(\hat{A}^+)=\overline{SW(\hat{A})}.$
\item
$SW(\hat{1})=1$ i.e. the constant function equal $1$ on the whole symplectic manifold ${\cal M}$ represents the identity operator.
\item
If an operator $\hat{A}$ is
self -- adjoint then $ SW(\hat{A})$ is a real function.
\item
The $\cdot\; $- product of operators is represented by the $*$- multiplication of functions, $SW(\hat{A}\cdot \hat{B})= SW(\hat{A})*SW(\hat{B}).$
\item
The trace of an operator equals
\be
\label{30}
{\rm Tr} \hat{A}= 
\frac{1}{(2 \pi \hbar)^n}\int_{\cal M}SW\big( \hat{A} \big) (q^1, \ldots, q^{2n}) t(q^1, \ldots, q^{2n})\omega^n. 
\ee
\end{enumerate}
The Stratonovich -- Weyl correspondence establishes a correspondence between a density operator $\hat{\varrho}$ and a Wigner function $W(q^1, \ldots, q^{2n})$ 
\be
\label{12.0}
SW\left( \frac{1}{(2 \pi \hbar)^n}\hat{\varrho}\right) = W(q^1, \ldots, q^{2n}).
\ee
Applying  the Stratonovich -- Weyl mapping to a density operator we find properties of a Wigner function on an arbitrary symplectic manifold.

Integration of a Wigner function yields
\[
 \int_{\cal M}W (q^1, \ldots, q^{2n}) t(q^1, \ldots, q^{2n})\omega^n=1. 
\]
 This result is an immediate consequence of the fact that ${\rm Tr}\hat{\varrho}=1.$
Moreover,
every Wigner function, as a counterpart of the self -- adjoint operator, is real.

Since a density operator is of a Hilbert -- Schmidt type, the integral satisfies the estimation
\[
\int_{\cal M}W^{*2} (q^1, \ldots, q^{2n})\, t(q^1, \ldots, q^{2n})\omega^n \leq \frac{1}{(2 \pi \hbar)^n}.
\]
The last equality holds only for pure states. The symbol $W^{*m} (q^1, \ldots, q^{2n}) $ denotes the $m$-th power of the function $W (q^1, \ldots, q^{2n})$ in the sense of the $*$- product. 

At the beginning of this paragraph we postulated (formula (\ref{10.5})) that observables are  formal series in the Planck constant $\hbar.$ On the contrary, there exist Wigner functions containing arbitrarily great negative powers of the deformation parameter i.e.
\be
\label{13}
W(q^1, \ldots, q^{2n})= \sum_{j= - \infty}^{\infty} \hbar^j W_j(q^1, \ldots, q^{2n})
\ee
and for arbitrary natural number $m$ there is such a natural number $s>m$ that 
\\ $W_{-s}(q^1, \ldots, q^{2n}) \neq 0.$ Indeed, from the Stratonovich -- Weyl correspondence and the fact that for a pure state its density operator is a projection operator, we see that for a pure state
\be
\label{16}
W^{*2}(q^1, \ldots, q^{2n})= \frac{1}{(2 \pi \hbar)^n} W(q^1, \ldots, q^{2n}).
\ee
Assume that $W(q^1, \ldots, q^{2n})= \sum_{j= 0}^{\infty} \hbar^j W_j(q^1, \ldots, q^{2n}). $ Thus
\[
W^{*2}(q^1, \ldots, q^{2n})= \sum_{i,j,k=0}^{\infty}\hbar^{i+j+k} B_k\big( W_i(q^1, \ldots, q^{2n}),W_j(q^1, \ldots, q^{2n}) \big).
\]
The local bilinear operators $B_k(\cdot, \cdot)$ define the $*$- product. 

Thus from the relation (\ref{16}) we see that $W_0(q^1, \ldots, q^{2n})=0.$ Using the mathematical induction we obtain that for every $j$ there must be $W_j(q^1, \ldots, q^{2n})=0$ so we arrive to a contradiction. Therefore at least Wigner functions of pure states must be of the form (\ref{13}).

Now we can show  why in general an integral form of a $*$- product is required. 
 If we calculate $W^{*2}(q^1, \ldots, q^{2n})$ as
\[
W^{*2}(q^1, \ldots, q^{2n})= \sum_{k=0}^{\infty}\sum_{i,j - \infty}^{\infty}\hbar^{i+j+k} B_k \big(W_i(q^1, \ldots, q^{2n}),W_j(q^1, \ldots, q^{2n}) \big),
\]
the term standing at $\hbar^m, \; m \in {\mathbb Z}$ is an infinite sum
\[
\sum_{i+j  \leq m} B_{m-i-j} \big(W_i(q^1, \ldots, q^{2n}),W_j(q^1, \ldots, q^{2n}) \big).
\]
Unless the number of nonzero operators $B_k(\cdot, \cdot)$ is finite, the sum above 
  can be divergent. Thus in calculations involving Wigner functions we have to apply another, for example  integral form of the $*$- product. Such a form is known for the Moyal product \cite{pleban}, \cite{gad}, \cite{var3}. 

\vspace{0.5cm}
The Stratonovich -- Weyl correspondence between the Hilbert space formulation and the phase space formulation suggests that criteria of being a Wigner function should be based on a notion of trace. Indeed, in contrary to the uniform convergence of a sequence of operators, which does not have an easy applicable counterpart in the phase space version of quantum mechanics, the trace is simply represented by the integral (\ref{30}). Therefore we can  transform Theorems \ref{tw5.0}, \ref{tw4.1} and \ref{co4.1} into their phase space versions.

We start from a counterpart of Theorem \ref{tw5.0}.

\begin{tw}
\label{tw5.0w}
A function   $W(q^1, \ldots, q^{2n}) $ defined on a symplectic manifold $({\cal M}, \omega)$ is a Wigner function iff:
\begin{enumerate}
\item
$\int_{\cal M}W^{*2} (q^1, \ldots, q^{2n})\, t(q^1, \ldots, q^{2n})\omega^n \leq \frac{1}{(2 \pi \hbar)^n},$
\item
the function is real,
\item
the series
\[
\frac{1}{2} \int_{\cal M}W^{*2} (q^1, \ldots, q^{2n})\, t(q^1, \ldots, q^{2n})\omega^n+
\sum_{l=2}^{\infty}\frac{(-1)^{l+1}}{l!} \frac{(2l-3)!!}{2^l} \times
\] 
\be
\label{6w}
\times
\sum_{r=0}^{l-1}(-1)^r \left( \begin{array}{c} l \\ r \end{array}\right) (2 \pi \hbar)^{2n(l-r-1)}\int_{\cal M}W^{*2(l-r)} (q^1, \ldots, q^{2n})\, t(q^1, \ldots, q^{2n})\omega^n =\frac{1}{(2 \pi \hbar)^n},
\ee
\item
$
\int_{\cal M}W (q^1, \ldots, q^{2n})\, t(q^1, \ldots, q^{2n})\omega^n = 1.$
\end{enumerate}
\end{tw}

An analog of Theorem \ref{tw4.1} can be formulated in the form
\begin{tw}
\label{tw4.1w}
A function   $W(q^1, \ldots, q^{2n}) $ defined on a symplectic manifold $({\cal M}, \omega)$ is a Wigner function of a quantum state iff:
\begin{enumerate}
\item
$\int_{\cal M}W^{*2} (q^1, \ldots, q^{2n})\, t(q^1, \ldots, q^{2n})\omega^n \leq \frac{1}{(2 \pi \hbar)^n},$
\item
the function is real,
\item
for every natural number $m$ the sum 
\be
\label{31}
\sum_{k=0}^m (-1)^k \left( \begin{array}{c} m \\ k \end{array} \right)(2 \pi \hbar)^{nk}\int_{\cal M}W^{*(k+1)} (q^1, \ldots, q^{2n})\, t(q^1, \ldots, q^{2n})\omega^n  \geq 0,
\ee
\item
$
\int_{\cal M}W (q^1, \ldots, q^{2n})\, t(q^1, \ldots, q^{2n})\omega^n = 1.$
\end{enumerate}
\end{tw}

Finally, a modified Theorem \ref{co4.1} reads as follows
\begin{tw}
\label{co4.1w}
A function   $W(q^1, \ldots, q^{2n}) $ defined on a symplectic manifold $({\cal M}, \omega)$ represents a quantum state iff:
\begin{enumerate}
\item
$\int_{\cal M}W^{*2} (q^1, \ldots, q^{2n})\, t(q^1, \ldots, q^{2n})\omega^n \leq \frac{1}{(2 \pi \hbar)^n},$
\item
the function is real,
\item
the limit 
\be
\label{32}
\lim_{m \rightarrow \infty} \; \sum_{k=0}^m (-1)^k \left( \begin{array}{c} m \\ k \end{array} \right) (2 \pi \hbar)^{nk}
\int_{\cal M}W^{*(k+1)} (q^1, \ldots, q^{2n})\, t(q^1, \ldots, q^{2n})\omega^n=0,
\ee
\item
$
\int_{\cal M}W (q^1, \ldots, q^{2n})\, t(q^1, \ldots, q^{2n})\omega^n = 1.$
\end{enumerate}
\end{tw}
Proofs of the presented criteria follow directly from their Hilbert space counterparts and the Stratonovich -- Weyl correspondence.

An identification method of pure states is based on the following statement.
\begin{tw}
\label{pw}
A function   $W(q^1, \ldots, q^{2n}) $ defined on a symplectic manifold $({\cal M}, \omega)$ represents a pure quantum state iff:
\begin{enumerate}
\item
$\int_{\cal M}W^{*2} (q^1, \ldots, q^{2n})\, t(q^1, \ldots, q^{2n})\omega^n = \frac{1}{(2 \pi \hbar)^n},$
\item
the function is real,
\item
$W^{*2} (q^1, \ldots, q^{2n})= \frac{1}{(2 \pi \hbar)^n}W(q^1, \ldots, q^{2n})$
\item
$
\int_{\cal M}W (q^1, \ldots, q^{2n})\, t(q^1, \ldots, q^{2n})\omega^n = 1.$
\end{enumerate}
\end{tw}
We emphasise that the main problem with testing Wigner functions is calculating their $*$-- product. As we said before, a differential form of this product can be useless in this case.

\underline{Example}

As an illustration of the presented criteria we examine a function considered  by Tatarskij in his distinguished paper \cite{tat}. The function
\be
\label{21}
W(p,q)= \frac{2}{3}W_0(p,q) + \frac{2}{3}W_1(p,q) - \frac{1}{3}W_2(p,q)
\ee
is defined on the phase space ${\mathbb R}^2.$ By $W_0(p,q), W_1(p,q)$ and $W_2(p,q)$ we denote the Wigner functions of mutually orthogonal states i.e.  
\[
\int_{{\mathbb R}^2}W_i(p,q)*W_j(p,q)dp dq= \frac{1}{2 \pi \hbar }\delta_{ij}.
\]  
The function $W(p,q)$ is not a Wigner function, because one of its eigenvalues is negative. Let us retrieve this observation by application of our three tests.

As it can be easily calculated 
\[
W^{*m}(p,q)= \frac{1}{(2 \pi \hbar )^{m-1}} \left(\left( \frac{2}{3}\right)^m W_0(p,q) + \left( \frac{2}{3}\right)^m W_1(p,q)+ \left(- \frac{1}{3}\right)^m W_2(p,q) \right).
\]
The function $W(p,q)$ is real, the integral $\int_{{\mathbb R}^2}W(p,q)dp dq=1.$ Moreover, \\ $\int_{{\mathbb R}^2}W^{*2}(p,q)dp dq=\frac{1}{2 \pi \hbar }$ so the function $W(p,q)$ satisfies the conditions $(i)$, $(ii)$ and $(iv)$ of Theorems \ref{tw5.0w}, \ref{tw4.1w} and \ref{co4.1w}. 

However,  for the function (\ref{21})  the sum (\ref{6w}) equals $\frac{5}{3} \cdot \frac{1}{2 \pi \hbar }$ so from Theorem \ref{tw5.0w} this function is not a Wigner function.

Now in Theorem \ref{tw4.1w} we see that the sum (\ref{31}) equals $2 \pi \hbar$ for $m=0,$ $0$ for $m=1$ and $-\frac{4}{9}\cdot 2 \pi \hbar$ for $m=2.$ Therefore  after taking three steps we conclude that $W(p,q)$ does not represent any state.

Finally, the limit (\ref{32}) in Criterion \ref{co4.1w} equals $- \infty$ so the tested function obviously cannot be a Wigner function.

After a slight modification this example can be considered on any $2$--D symplectic manifold. 

{\bf Acknowledgments}
 
This work was supported by the CONACYT (Mexico) grant No. 103478. We are grateful to Prof. Maciej Przanowski and Dr. Michal Wasiak for their interest in this paper and valuable remarks. 

%%%%%%%%%%%%%%%%%%%%%%%%%%%%%%%%%%%%%%%%%%%%%%%%%%%%%%%%%%%%%%%%%%
%%%%%%%%%%%%%%%%%%%%%%%%%%%%%%%%%%%%%%%%%%%%%%%%%%%%%%%%%%%%%%%%%

\end{document}